\documentclass[manuscript,authoryear,12pt]{elsarticle}
\usepackage{graphicx}
\usepackage{natbib}
\usepackage{amssymb}
\journal{Journal of Theoretical Biology}

\usepackage[nofiglist,notablist]{endfloat}

\begin{document}

\begin{frontmatter}

\title{Anomalies in the transcriptional regulatory network of the yeast {\it
    Saccharomyces cerevisiae}}

\author{M. Tu\u{g}rul$^{\dagger\,*}$ and A. Kabak\c{c}{\i}o\u{g}lu$^\dagger$}
\address{$^\dagger$ Department of Physics, Ko\c{c} University, Sar{\i}yer 34450 Istanbul, Turkey\\ $^*$ IFISC(CSIC-UIB) Institute for Cross-Disciplinary Physics and Complex Systems, Campus Universitat Illes Balears, E-07122 Palma de Mallorca, Spain}

\begin{abstract}
  We investigate the structural and dynamical properties of the
  transcriptional regulatory network of the yeast {\it Saccharomyces
    cerevisiae} and compare it with two ``unbiased'' ensembles: one
  obtained by reshuffling the edges and the other generated by
  mimicking the transcriptional regulation mechanism within the
  cell. Both ensembles reproduce the degree distributions (the first
  -by construction- exactly and the second approximately),
  degree-degree correlations and the $k$-core structure observed in
  Yeast. An exceptionally large dynamically relevant core network
  found in Yeast in comparison with the second ensemble points to a
  strong bias towards a collective organization which is achieved by
  subtle modifications in the network's degree distributions. We use a
  Boolean model of regulatory dynamics with various classes of update
  functions to represent in vivo regulatory interactions. We find that
  the Yeast's core network has a qualitatively different behaviour,
  accommodating on average multiple attractors unlike typical members
  of both reference ensembles which converge to a single dominant
  attractor. Finally, we investigate the robustness of the networks
  and find that the stability depends strongly on the used function
  class. The robustness measure is squeezed into a narrower band
  around the order-chaos boundary when Boolean inputs are required to
  be nonredundant on each node. However, the difference between the
  reference models and the Yeast's core is marginal, suggesting that
  the dynamically stable network elements are located mostly on the
  peripherals of the regulatory network. Consistently, the
  statistically significant three-node motifs in the dynamical core of
  Yeast turn out to be different from and less stable than those found
  in the full transcriptional regulatory network.

\end{abstract}

\begin{keyword}
Boolean dynamics \sep attractor distribution \sep robustness \sep dynamical core network \sep canalizing functions
\PACS 89.75.-k \sep 64.60.Fr \sep 36.20.Ey

\end{keyword}

\end{frontmatter}

\section{Introduction}
Transcriptional regulatory network (TRN) describes the connective structure of
the gene-gene interactions that regulate most physiochemical activities in a
cell~\citep{Thomas_LawsDynamicsRN,AlbertBarabasi_StatMecCompNets}. This is
particularly valid for {\it Saccharomyces cerevisiae} (from here on referred
to as ``Yeast''), an eukaryote which lacks miRNA and RNAi capability. Survival
of the cell under changing external conditions requires both a responsive and
a robust regulatory mechanism~\citep{Kauffman_Origins_of_Order}. The two key
ingredients that contribute to the dynamics of regulatory activity are the
network's topology and the character of the regulatory
interactions~\citep{NCF}. It is natural to attempt to identify the respective
roles of these two components on the dynamical properties of the system. For
this purpose, we consider here two neutral network ensembles and compare their
representative members with the yeast's transcriptional regulation
network. The first ensemble (E1) is generated by reshuffling the edges on the
Yeast's TRN. This ensemble is suitable for identifying features that fall
beyond those implied by the node connectivities. The second ensemble (E2) was
developed in a recent study by \cite{BKME} and was shown to mimic with high
accuracy the global structural properties of the network of transcriptional
regulatory interactions~\citep{LeeYeast,yeastract} found in Yeast. By
comparing the characteristic features of the regulatory dynamics on Yeast's
TRN with those of the two ensembles, we discuss the relevance of in/out-degree
statistics and the functional character of the interactions to the regulatory
dynamics.

Below, we first focus on the structure of the TRN found in Yeast and those
of the typical members of the two ensembles. Next, we investigate the
differences between their dynamics in terms of their attractor
statistics and robustness, both within the framework of a synchronous,
Boolean time-evolution model.

\section{Structure of Yeast's TRN}

A reasonably complete picture of the architecture of the TRN in Yeast is now
available due to the collective effort of many experimental groups and the
recent development of high-throughput
techniques~\citep{Spellman_Yeast,LeeYeast,yeastract}. The structure is radically
different from a random collection of $4252$ nodes connected by $12541$ edges,
where the quoted numbers are those of the genes and known regulatory
interactions in the Yeastract database~\citep{yeastract} adopted
by \cite{BKME}. In particular, the regulating nodes (transcription factors,
or TFs) which constitute $\sim 3.5\%$ of all genes have a skewed out-degree
distribution which has been suggested to follow a
power-law~\citep{Guelzim_Yeast,Bergman_etal_SixGRN,Maslov_Sneppen}, and a
roughly exponential in-degree distribution~\citep{Maslov_Sneppen}, although
the ranges of both distributions are somewhat narrow to make a strong
case. Nevertheless, the deviation from a randomly wired network is strong and
independent of the database used. A similar trend is observed in other
structural aspects, such as degree-degree correlations and the $k$-core
organization which appear to follow from the in- and out-degree
distributions~\citep{BKME}. Note that the numbers quoted above continue to
increase as new experimental data pours in.  However, for the sake of a fair
comparison with the earlier study we stick to the same reference data and pay
attention that our conclusions are based on observations (such as normalized
distributions) that are likely to change little with further incoming data.

\subsection{Reference ensembles}

The ensemble E1 is an ``unbiased'' set of networks generated from Yeast's TRN
by shuffling the edges among the nodes while keeping the in- and out-degrees
of each node unaltered. This is achieved simply by switching the end terminals
of two randomly picked edges $(i\rightarrow j),(k\rightarrow l)$ to obtain two
new edges $(i\rightarrow l),(k\rightarrow j)$, where we impose $i\neq j\neq
k\neq l$ in order to keep the number of self-regulating genes fixed (see
below). In spite of its frequent use in the literature, one should consider E1
a biologically inappropriate reference point. There is no good reason to
assume that Nature selected Yeast's TRN out of a pool of networks with
identical degree distributions. Furthermore, such a reference simply ignores
the question pertaining to the origin of the observed degree distribution
for which several mechanisms have been
proposed~\citep{wagner1994egn,vannoort2004ycn,BKME}.  Nevertheless,
stastically significant deviations from E1 found in Yeast (features with a
high z-score) allowed past studies to point out a high abundance of stability
enhancing local motifs, such as feed-forward loops. On the other hand, the
highest z-score 3-node motif thus found by \cite{Prill_Motif} can be reproduced
with the right frequency by means of a simple model which we also consider
here to generate a second reference ensemble E2 as descibed below.

E2 is generated from a biologically motivated model introduced by
\cite{BKME}. In this null model, two binary strings are associated
with each yeast gene: first ($S_1$) representing the promoter site of
the gene that regulates its transcription, second ($S_2$) representing
the DNA sequence (motif) that the gene's product binds. The lengths of
the two sequences are chosen randomly from the associated length
distributions determined from the available biological data provided
by \cite{Harbison_RSdistr}. A gene $A$ is said to regulate gene $B$ if
$S^A_2 \subset S^B_1$. No such $B$ exists unless the product of $A$ is
a TF. For more details, we refer the reader to the original article.

Repeated generation of model networks with the same number of genes as
in Yeast and different random number generator seeds forms the
ensemble E2. Its typical members agree to very good accuracy with
Yeast in terms of the in-, out-, and total-degree distributions,
degree-degree correlation and the rich-club coefficient
distributions. This ensemble further captures the hierarchical
organization of Yeast's TRN given by its $k$-core analysis, as well as
the frequencies of most 3-node motifs observed in Yeast (see
Section~\ref{motifs_sec}). Due to its success in simultaneously
reproducing several distributional features of Yeast's TRN from a
single mechanism and {\it without any reference to the genetic
  sequence}, E2 appears a meaningful alternative to E1. It is a
null-hypothesis constructed by a bottom-up approach, providing a
microscopic explanation for a number of ``anomalous'' distributions
observed in Yeast's TRN and taken for granted in randomized ensembles
such as E1. However, as shown below, even a quantitative agreement on
those frequently quoted features may veil some significant differences
between the actual Yeast and the two ensembles.

\section{Dynamics}
\subsection{Boolean network model}
Several methods can be employed for simulating the time evolution of
gene expression within a cell~\citep{Norell_etal_ContAttractor}. We
here use a Boolean network model first proposed by
\cite{Kauffman_Network}, where the expression level $\sigma_i(t)$ of
the $i^{th}$ gene at time $t$ is assumed $0$ (silent) or $1$
(expresed). The interaction between the regulatory genes and the
regulated gene is deterministic and the time evolution is synchronous,
so that
\[
\sigma_i(t+1) =
F_i(\sigma_{j_1}(t),\sigma_{j_2}(t),\dots,\sigma_{j_{d_i}}(t))\ ,
\]
where $d_i$ is the number of edges incoming to the node $i$ (number of TFs
regulating the $i^{th}$ gene), $\{j_1,\dots,j_{d_i}\}$ are the nodes connected
to $i$ with incoming edges, and $F_i$ is a Boolean function determining
the state of gene $i$ in presence of possibly multiple regulators. 

Boolean dynamics of generic (such as, fixed
in-degree~\citep{Kauffman_Network},
random~\citep{AldanaCluzel_Random},
power-law~\citep{Aldana_BooleanNetPLtopology2003}) networks have been
of considerable interest for some time. Applications to biological
systems include TRN models of {\it Arabidopsis thaliana} by
\cite{Mendoza_etal_Arabidopsis}, {\it Drosophila melanogaster} by
\cite{AlbertOthemer_TopologyPredictsExpression}, and the Yeast's
cell-cycle network studied by \cite{Li_etal_YeastRobustness}.  Unlike
most past work which focused on part of the TRN associated with a
particular function, we here focus on the {\it global} dynamical
behavior of yeast and yeast-like regulatory networks. A recent work by
~\citep{Rieger} in the same spirit compares Yeast and {\it E. coli} on
a full-scale.

Determining the time evolution of the network starting from a given
initial state is straightforward once the regulatory functions
$\{F_i\}$ are fixed. However, the nature of these interactions in the
actual organism are too complex and far from being well
understood. Therefore, instead of attempting to identify the functions
$F_i$ that best describes the behavior of the yeast cell, we
investigated the generic dynamical properties of the known
architecture arising from a random choice of $\{F_i\}$ picked from a
suitable collection.

It has been suggested that $\{F_i\}$ may be further restricted, based
on the available experimental evidence, to certain subclasses of
Boolean functions~\citep{Kauffman_Origins_of_Order,NCF,SNCF}. These
are

{\it Simple Random Functions (RF)}: Each input is randomly assigned an
output value of $1$ with a probability $p$ and $0$ otherwise.

{\it Canalizing Functions (CF)}: It has at least one canalizing input, say
$\sigma_{j_1}$, such that $F_i(\sigma_{j_1}=\zeta,\dots)=s$.  $\zeta$ and
$s$ are called the canalizing value and the canalized output,
respectively~\citep{Kauffman_Origins_of_Order}. For consistency, we set
$s=1$ with probability $p$ and let $F_i(\sigma_{j_1}=\bar{\zeta},..)$
be a simple random function of the remaining variables.

{\it Nested Canalizing Functions (NCF)}: Also called Hierarchical
Canalizing Functions, they were proposed by \cite{NCF} based on an
earlier analysis due to \cite{Harris} and claimed to more closely
mimic actual biological systems. All input variables of a NCF are
canalizing and ordered in rank. The canalized output is determined by
the highest ranking node at its canalizing value and is chosen to be
$1$ with probability $p$.

{\it Special Nested Canalizing Functions (SNCF)}: Recently, gene
regulation dynamics was observed to be consistent with a more restricted
subclass of NCFs given by the Boolean expressions
\begin{eqnarray}
\label{SNCF60}
&&\tilde{\sigma}_{j_1} \wedge (\tilde{\sigma}_{j_2} \wedge (... \wedge
(\tilde{\sigma}_{j_{d_i}-1} \wedge \tilde{\sigma}_{j_{d_i}})...)) \nonumber \\
\label{SNCF30}
\mbox{and}\ &&\tilde{\sigma}_{j_1} \wedge (\tilde{\sigma}_{j_2} \wedge (... \wedge
(\tilde{\sigma}_{j_{d_i}-1} \vee \tilde{\sigma}_{j_{d_i}})...))
\end{eqnarray}
where $\tilde{\sigma} \in \{\sigma,\bar{\sigma}\}$ and the probability of
occurance of the two functions were found to be $0.66$ and $0.29$,
respectively~\citep{SNCF}. The function classes above satisfy SNCF $\subset$
NCF $\subset$ CF $\subset$ RF.

A constraint on these functions is that each input variable should be
relevant, so that they have an experimentally detectable regulatory
signature. Accordingly, for each input $\sigma_{j_i}$ of a function $F_i$ with
$d_i$ inputs, there exists at least one input set
$\{\sigma_{j_1},..,\sigma_{j_{d_i}}\}$, such that,
\begin{equation}
\label{relevant}
F_i(\sigma_{j_1},..,\sigma_{j_i},..,\sigma_{j_{d_i}}) \ne
F_i(\sigma_{j_1},..,\bar{\sigma}_{j_i},..,\sigma_{j_{d_i}})
\end{equation}
with $\bar{\sigma} \equiv 1-\sigma$. We ensure that the condition
in Eq.~(\ref{relevant}) is satisfied in all cases.

Let $\pi$ be the fraction of inputs to $F_i$ that satisfy Eq.~(\ref{relevant}) for
which the output equals $1$ calculated over all genes:
\begin{eqnarray}
\pi &=& \big(1/N\big)\sum_i \delta_{F_i(\sigma_{i_1},..,\sigma_{i_{d_i}}),1} /
2^{d_i}
\end{eqnarray}
with $N = $ total number of genes and $\delta_{i,j}$ is the Kronecker
delta function. For the SNCF class, $\pi$ is not a free parameter but
determined by the in-degree distribution of the network. For the rest
of the function types we have $\pi \neq p$ in general, because
Eq.~(\ref{relevant}) filters out some function assignments. In fact,
the fraction $\phi_n$ of the Boolean functions satisfying
Eq.~(\ref{relevant}) with $n$ inputs is
well-known~\citep{nondeg_boolean}:
\begin{eqnarray}
\label{nondegenerate}
\phi_1&=&2/4 \nonumber \\
\phi_2&=&10/16 \nonumber \\
\phi_3&=&218/256 \nonumber \\
\phi_4&=&64594/65536
\end{eqnarray}
where the denominators are $2^{2^n}$: number of Boolean functions with
$n$ inputs. For a fair comparison (except in the discussion on
robustness), we adjusted the parameter $p$ for RF, CF, and NCF, so that
$\pi \simeq 0.29 $, the value one obtains both for Yeast and the two
ensembles with a SNCF assigned to each gene.

We will compare the impact of each function class on the system's dynamics
when implemented on the dynamically relevant core network of the TRN described
in the next section.

\subsection{Dynamically Relevant Core}

The number of the dynamical attractors in a Boolean network is determined by a
{\it dynamically relevant core} network (DRC) involving TFs only. DRC in Yeast
is much smaller than the full transcriptional regulatory network and allows
one to perform time-efficient simulations of the regulatory dynamics.

\begin{figure}[h!]
\vspace*{0.5cm}
  \begin{center}
    \includegraphics[width=3in]{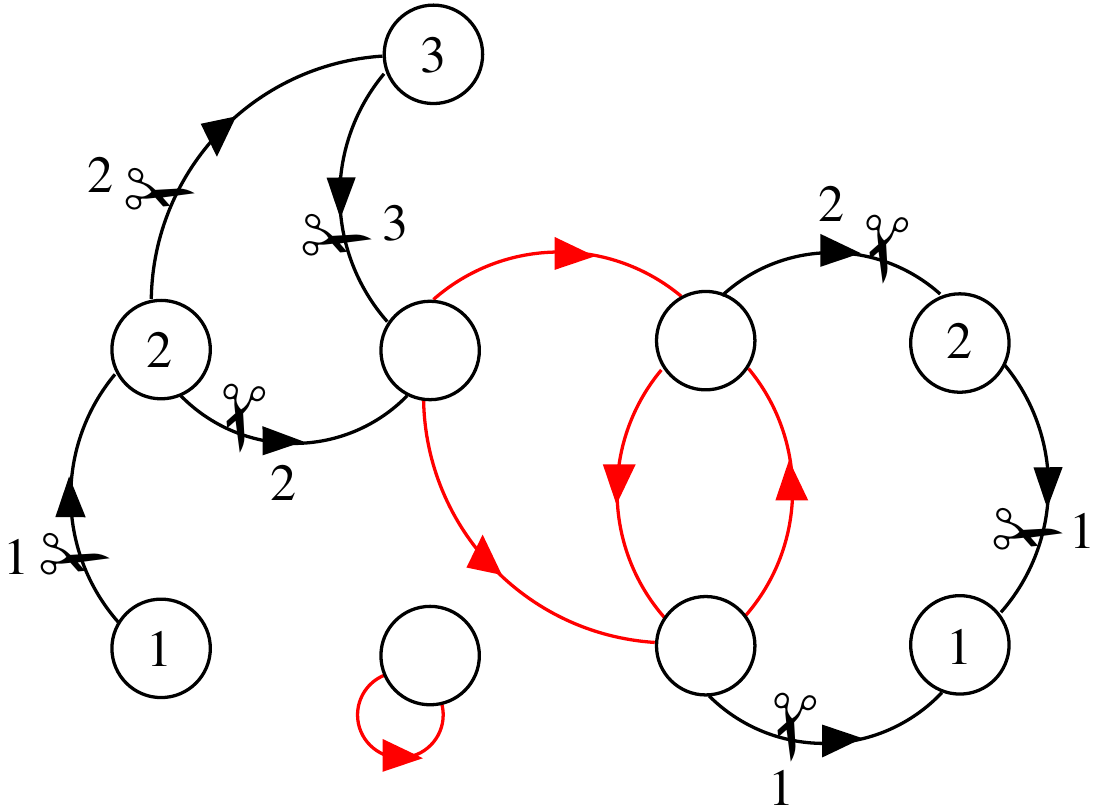}
  \end{center}
\caption{}
\label{drs_prune}
\end{figure}

We define a global DRC for Yeast by recursively pruning all the nodes with
zero in-degree or zero out-degree (see Fig.~\ref{drs_prune}). Most of the
genes in the transcriptional regulatory network of Yeast are ``slave'' genes
that do not take part in the regulation process or do so merely in a
downstream fashion. Their presence may effect the transient behavior or the
size of the dynamical cycles the network eventually falls into, however they
do not change the number of distinct steady-states (or attractors, see below)
and their probabilities of occurance under different function
assignments. These genes can be pruned by recursively eliminating nodes with
zero out-degree.

In a similar fashion there exist a set of genes with zero in-degree, whose
expression levels remain unchanged throughout the dynamics. Some of these are
'housekeeping' genes that are always expressed to perform routine functions
(such as RNA production) while others are fixed by the environmental
conditions. We will consider the regulatory dynamics when the states of these
genes are fixed and prune them as well, assuming that the averaging over
different state assignments to such genes is properly accounted for by
different function assignments on the remaining network.
See the recent work by \cite{Rieger} for an earlier implementation of
this procedure on Yeast, while for a different reductionist approach
see \cite{PaulKaufmanDrossel_RelevantNodes}.
\begin{figure}[h!]
\vspace*{0.5cm}
  \begin{center}
    \includegraphics[width=5in]{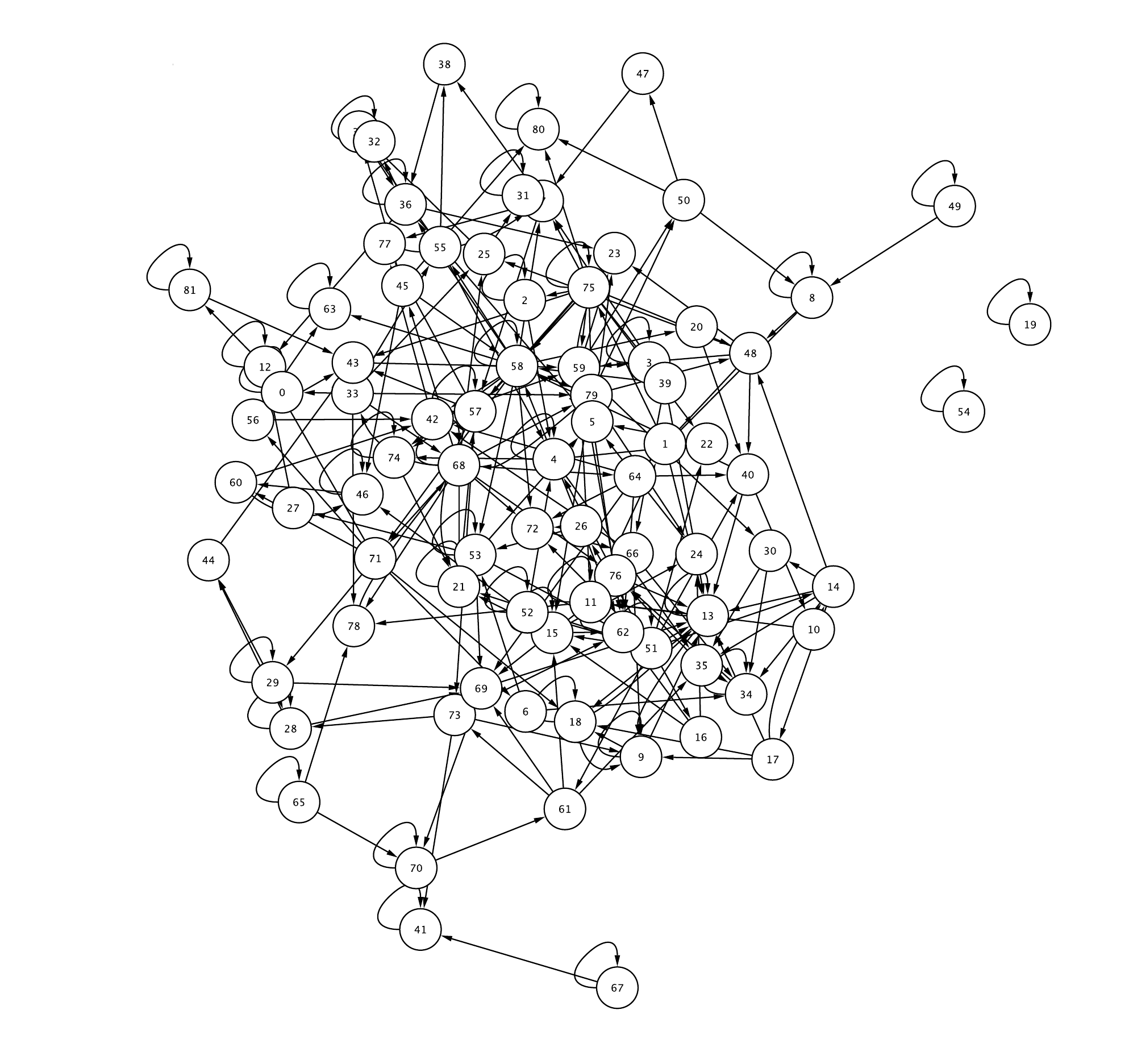}
  \end{center}
\caption{}
\label{yeast_trn}
\end{figure}

The end product (shown in Fig.~\ref{yeast_trn} for Yeast) is a subnetwork
where each node is a TF regulated by other genes in the DRC and/or by
itself. Given that only $3.5\%$ of all the genes in {\it Saccharomyces
  cerevisiae} are TFs, described pruning process brings a sizeable reduction
in the computation time. We found that Yeast's dynamically relevant subnetwork
contains $82$ TF genes and $254$ interactions between them.

\subsubsection*{Comparison of DRC sizes in Yeast {\it vs} reference ensembles}

\begin{figure}[h!]
\vspace*{0.5cm}
  \begin{center}
    \includegraphics[width=5in]{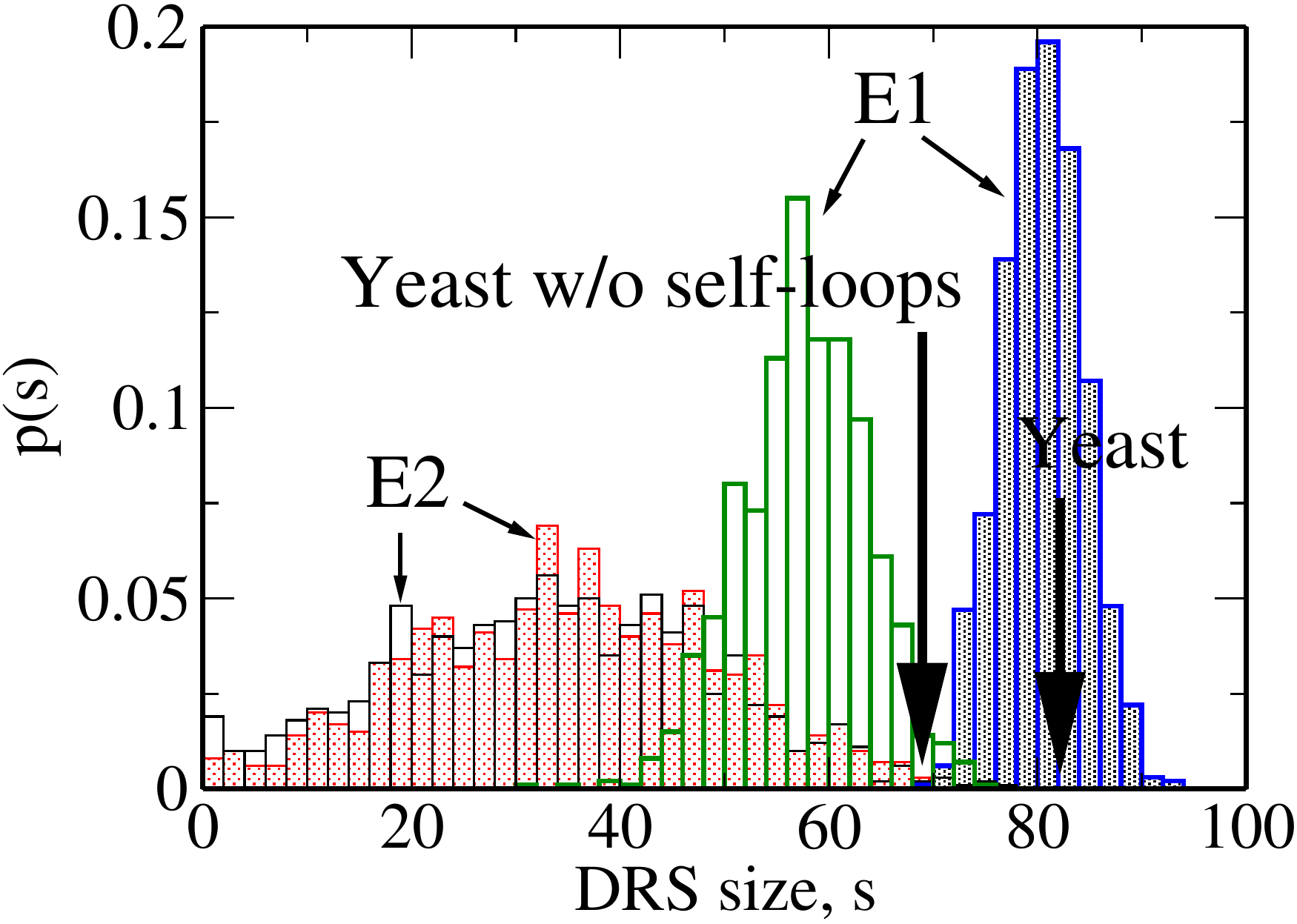}
  \end{center}
\caption{}
\label{drc_size}
\end{figure}

We also obtained the DRC for $1000$ networks chosen from each ensemble. In the
case of E2, ensuring that the samples before pruning have, on average, the
same number of nodes and edges as Yeast's TRN . The comparison of the three
cases is shown in Fig.~\ref{drc_size}. The outcome is instructive: The
distribution of the DRC size on E2 shows a big contrast with Yeast while the
agreement with the edge-reshuffled ensemble E1 (with {\it exactly} the same
in- and out-degrees at each node) is perfect when self-regulating genes are
included and still better than E2 otherwise. (See next section for a
discussion on self-regulating genes.) A similar situation is observed also for
the number of interactions in the DRC. In view of the quantitative agreement
of in- and out- degree-distributions between Yeast and E2~\citep{BKME}, this
result points to a subtle difference between the two ensembles that translates
into a three-fold size difference in the respective DRCs.

There is no doubt that, a complete model of the regulatory dynamics
should include the environmental inputs carried by the signaling
pathways and the post-transcriptional regulatory
interactions~\citep{Thomas_LawsDynamicsRN,SamalJain}. Accordingly, a
complete DRC should include of both TFs and non-TF proteins (see,
e.g. \cite{Li_etal_YeastRobustness}). Nevertheless, the
``transcriptional regulatory'' DRC defined as above remains an
intrinsic property of the organism, which in the case of Yeast appears
to be significantly out of proportion.

\subsection{Number of Attractors}

The dynamical characterization of a TRN is complete once all the interactions
(the network's architecture) and the update functions $\{F_i\}$ of all genes
are specified. The network state $S(t) \equiv
\{\sigma_1(t),\sigma_2(t),\dots,\sigma_N(t)\}$ at the discretized time
$t=n\Delta t$ after $n$ update cycles can be expressed symbolically as
\begin{eqnarray}
S(t) &=& F^n(S(0)) \ .
\end{eqnarray}
The time unit $\Delta t$ is assumed to be large enough to encompass all
protein production related processes. Since the time evolution is
deterministic, the number of possible dynamical trajectories is $2^N$, i.e.,
the number of distinct initial conditions.  $2^N$ is also the size of the
state space, therefore each initial condition eventually ends up in a cycle
which is called the $attractor$ for that initial state. A state $S_k$ is a
member of the attractor if and only if $S(t+n\Delta t)=S(t)=S_k$ for some
integer $n > 0$. Minimal such $n$ is called the $length$ of the attractor
cycle. The attractor is a {\it fixed point} if $n=1$ and a {\it limit-cycle}
otherwise. Note that the attractor lengths (except those of the fixed points)
are modified after the pruning process describe above, but the number of
attractors is not. Therefore we focus on the statistics of their number below.

We note in passing that, the attractors of a transcriptional regulation
network may be associated with the observable features of the organism. For
example, the expression pattern of the segment polarity genes in {\it
  Drosophila melanogaster} can be mapped to the fixed point of the relevant
regulatory subnetwork~\citep{AlbertOthemer_TopologyPredictsExpression}. In
{\it Arabidopsis thaliana}, the attractors of the regulatory subnetwork
responsible from cell differentiation have been shown to correspond to
different phenotypes~\citep{Mendoza_etal_Arabidopsis}.

\begin{figure}[h!]
\vspace*{0.8cm}
  \begin{center}
    \includegraphics[width=4in]{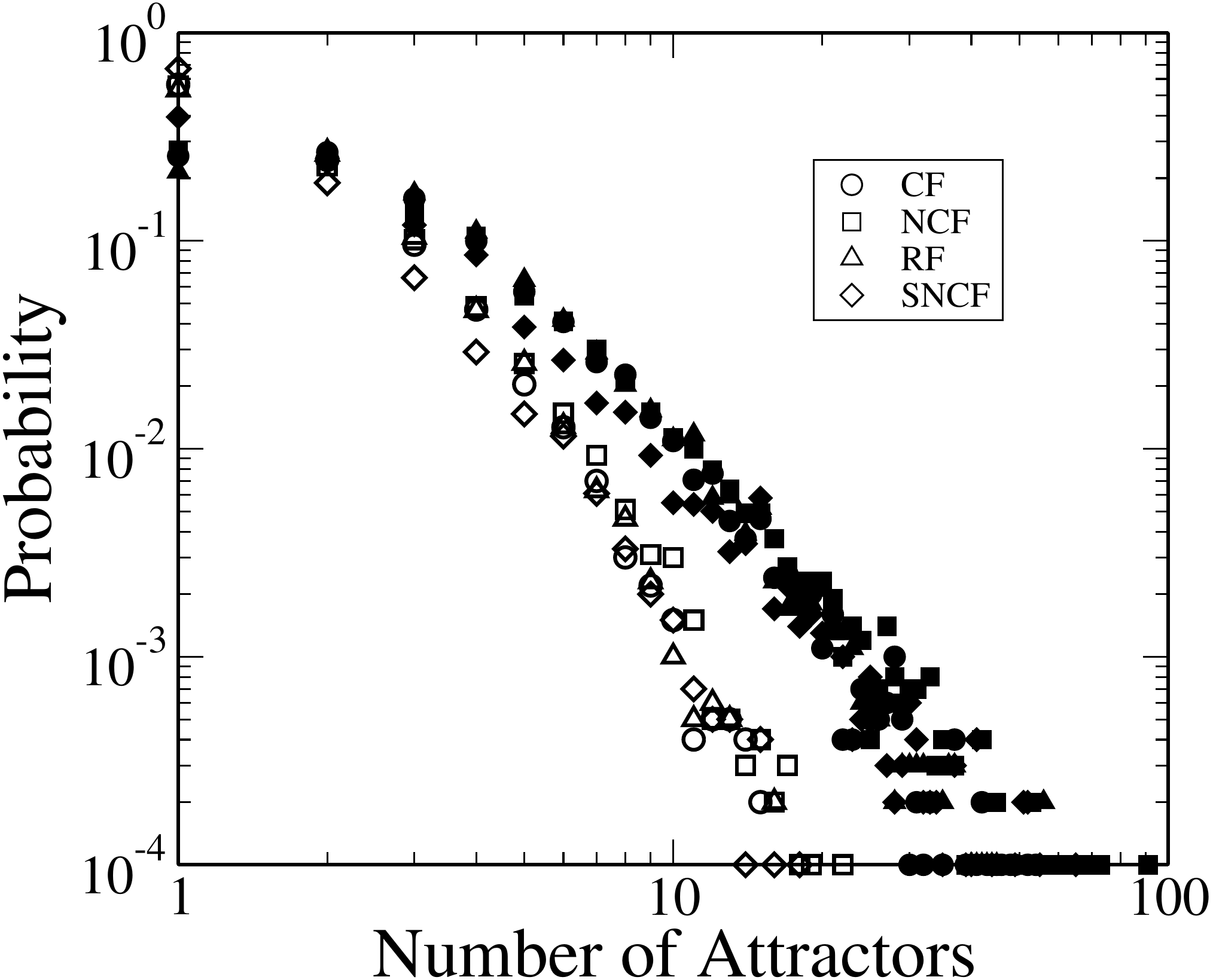}
  \end{center}
\caption{}
  \label{n_distr}
\end{figure}

The number of attractors, $N_{att}$, remains invariant as we switch from the
full-size network to the DRC, but estimating its exact value is difficult due
to the vast number of initial conditions that should be checked. However, by
randomly sampling a small number of initial conditions, another - and possibly
more relevant - measure of the number of attractors, $\tilde{N}_{att}$, can be
easily calculated. $\tilde{N}_{att}$ is obtained by properly weighting the
$i^{th}$ attractor by its {\it basin of attraction} $\omega_i$, the number of
initial conditions that end up in $i$. This procedure allows one to
distinguish between, e.g. two network realizations (network + function
assignments) with two attractors each, whose relative basin sizes are $(0.99,
0.01)$ in the first case and $(0.5, 0.5)$ in the second. It is fair to say
that the first network realization has a single dominant attractor, while the
second one has two. The generalization of this argument gives $\tilde{N}_{att}
= 2^S$, where $S$ is the standard dynamical entropy of the network (see, e.g.,
\cite{Kravitz_Shmulevich}):
\begin{eqnarray}
\label{entropy}
S = -\sum_i p_i\log_2p_i
\end{eqnarray}
where $p_i=\omega_i/2^N$ is the probability that a uniformly selected initial
condition is in the basin of attraction of the $i^{th}$ attractor. Note that
$\tilde{N}_{att} \le N_{att}$. The difference between the two attractor counts
is demonstrated in Fig.~\ref{n_distr} over the E2 ensemble.

We estimated the average number of attractors for Yeast's DRC and the model
DRCs. Since an exact enumeration is out of question with $2^{82}$ possible
initial states (and unnecessary for estimating $\tilde{N}_{att}$), we randomly
sampled $1000$ initial conditions and followed their trajectories in time until
an attractor was reached. The attractor was then characterized by the sequence
of states in the cycle (or by the state id for fixed points) and the number of
initial conditions that end up in each state were counted.

$\tilde{N}_{att}$ thus found for the networks were further averaged
over $1000$ different structures, with each structure analyzed using
$10$ independent Boolean function assignments chosen from each of the
classes above. Mean attractor numbers thus found for the E2 ensemble
are $2.3$, $2.2$, $2.3$ and $1.9$ when the update functions were
chosen from the classes RF, CF, NCF and SNCF, respectively. We
deliberately omitted the error bars in the values reported above,
because the distribution of the attractor number is highly skewed and
has a fat tail, as shown in Fig.~\ref{n_distr}. A fair comparison
with Yeast and E1 requires some more work and is presented
below. However, except for a somewhat smaller average attractor count
observed for SNCFs, neither on Yeast nor on the reference ensembles
did we find a significant difference among the attractor statistics of
different function classes.

\subsection{Robustness of the dynamics}

The survival of a cell relies on the continuously and reliable
production of a vast amount of proteins in proper
quantities. Therefore, the ``equilibrium'' gene expression profile
(encoded here by an attractor) is expected to possess a certain level
of stability, i.e., robustness to perturbations such as random
fluctuations in expression levels or temporary malfunction of a
gene. On the other hand, a certain level of responsiveness is also
necessary in order to be able to cope with the environmental changes
in longer time scales. This trade-off suggests that living organisms
function in the vicinity of the order-chaos boundary, a hypothesis
originally formulated by \cite{Kauffman_Origins_of_Order}.

A dynamical system is said to be chaotic if a small perturbation introduced
into one of its two, otherwise identical, copies drives them away from each
other exponentially fast. Adopted to TRN dynamics, this amounts to monitoring
in time the normalized ``Hamming distance'' 
\[ h(S,S') = (1/N)\sum_i^N[\sigma_i-\sigma_i']\]
between the two copies $S$ and $S'$.  The
network's robustness is then determined by
\begin{equation}
r=\lim_{h\rightarrow0^+, t\rightarrow \infty}\big\langle dh(t+\Delta
  t) / dh(t) \big\rangle\ ,
\label{r_eq}
\end{equation}
where the first limit ensures that the measured quantity is a
steady-state property (a function of the attractors only), the second
reflects that $r$ is a linear response function. $\Delta t$ is chosen
to be a small time interval (one time step in our case) and the
averaging is over possible
perturbations~\citep{Aldana_BooleanNetPLtopology2003}. The network is
said to be chaotic if $r>1$ and ordered if $r<1$ (both in an average
sense, for there may be particular perturbations in each case that
result in the opposite behavior.) We will show below that, RF, CF, NCF
and SNCF display significant variability in robustness on the Yeast and
the model networks.

\section{Yeast {\it vs} model networks}

\subsection{Attractor statistics}
Performing the same analysis on Yeast's TRN and E1, one finds that the average
number of attractors is orders of magnitude larger for RF type functions. In
fact, most of the initial conditions end up in different attractors, so that
$\tilde{N}_{att}$ is capped by the number of initial conditions used for
averaging. Such disagreement is too large to be explained by the mismatch in
DRC sizes shown in Fig.~\ref{drc_size}. For an estimate of the network-size
dependence of $N_{att}$ see, e.g., \cite{Drossel_Natt} and \cite{ErzanBalcan}.

Upon closer inspection, the excessive attractor number turns out to be due to
another structural anomaly in Yeast: the presence of a large number of
self-regulating genes.  $16\%$ of all genes in the DRC ($13$ out of $82$) are
self-regulating~\citep{yeastract} as opposed to $5\%$ (less than $2$ out of an
average of $35.6$ - see Fig.~\ref{drc_size}) in E2 model. Note that a
self-regulating gene is by definition a member of the DRC and is preserved
under the edge shuffling process used for generating E1 networks.  Each such
node under synchronous Boolean dynamics potentially doubles the attractor
count due to a parity effect. The excess of such nodes in Yeast's TRN is well
known (see, e.g., \citep{Rieger}), although it can easily be missed in a
structural comparison as in \cite{BKME} unless specifically looked for. Its
dynamical signature, however, is difficult to overlook. This example serves as
a demonstration for how a comparative study of the dynamics may lead to the
discovery of structural features specific to Yeast.


We next eliminate the self-loops from Yeast and both model networks, and
reconstruct the DRCs. By doing so, we temporarily depart from a faithful
representation of Yeast. On the other hand, we ensure that further
discrepancies we may encounter originate from structural differences other
than the high frequency of self-regulating genes in Yeast. The
self-interactions will be restored in the next section.
\begin{table}
\begin{center}
\begin{tabular}{lcccc}
\hline
&RF & CF & NCF & SNCF \\
\hline\hline
E1 & 1.9 & 1.8 & 1.8 & 1.7\\
\hline
E2 & 1.6 & 1.6 & 1.6 & 1.5\\
\hline
E2$^*$ & 1.7 & 1.6 & 1.7 & 1.6\\
\hline
Yeast & 4.8 & 4.2 & 4.1 & 3.4\\
\hline
\end{tabular}
\caption{}
\label{N_avg}
\end{center}
\end{table}

\begin{figure}[h!]
\vspace*{1.2cm}
  \begin{center}
    \includegraphics[width=4in]{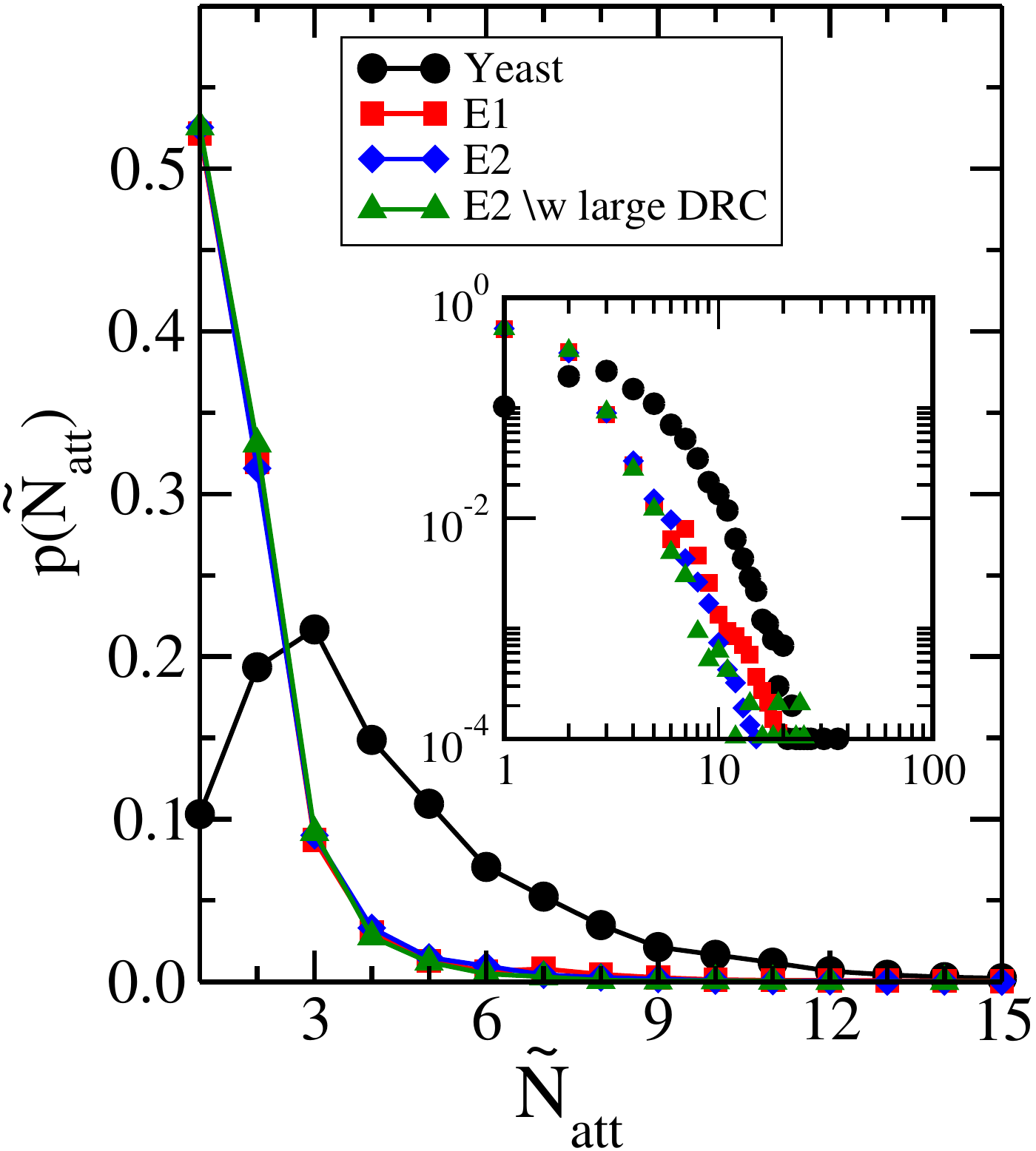}
  \end{center}
\caption{}
\label{noloop_n_distr}
\end{figure}

Interestingly, the histograms $p(\tilde{N}_{att})$ obtained from Yeast now
differ significantly from those of both reference
ensembles. Fig.~\ref{noloop_n_distr} displays the contrast for RF type
functions, while qualitatively the same picture is obtained also for other
function classes. Majority of E1 and E2 model networks are dominated by a single
attractor, while Yeast's DRC typically has multiple attractors and and a
nonmonotonous $p(\tilde{N}_{att})$. The contrast between the average attractor
numbers is shown in Table~\ref{N_avg}. The tail of the attractor number
distribution obtained from Yeast is also markedly different, as shown in the
inset of Fig.~\ref{noloop_n_distr}.

It is not possible to understand this dynamical anomaly of Yeast in terms of the
difference in the DRC sizes shown in Fig.~\ref{drc_size}. Because, first, the
mismatch in $\tilde{N}_{att}$ persists between E1 and Yeast which have
identical DRC sizes. Second, although a smaller ``random'' network is expected
to have a smaller number of attractors on average
\citep{Drossel_Natt,ErzanBalcan}, we observed above that E1 and E2 networks
(without self-regulating nodes) which have approximately a factor of two size
difference come with almost the same number of attractors. We further
confirmed this second observation by repeating our analysis on a filtered
subset of E2 networks (again, without self-loops) whose number of TFs and the
number of interactions in the DRC are within $5\%$ of Yeast's. $\tilde{N}_{att}$
is essentially the same as before (also shown in Fig.~\ref{noloop_n_distr}). Therefore, we conclude that \\
1. Synchronous Boolean dynamics on DRC-type networks has an attractor
statistics with a much weaker size dependence than random networks,\\
2. Yeast's TRN has certain structural elements 
that amount to a significant modification of its attractor statistics and
are not captured by either of the two model ensembles E1 \& E2.

\subsection{Robustness}

\begin{figure}[h]
\vspace*{0.8cm}
\begin{center}
\includegraphics[width=4in, angle = 0]{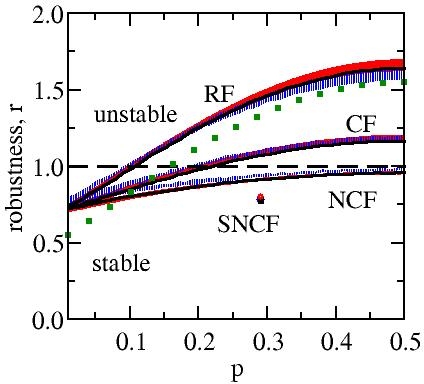}
\end{center}
\caption{}
\label{YeastRobustness}
\end{figure}

We performed the analysis outlined above on Yeast and also on E1 and
E2 networks with similar DRC sizes (and with self-interactions
untouched) as described earlier. We found that, the robustness depends
significantly on the type of the update function: the more restrictive
the function set, the more robust is the dynamics (see
Fig.~\ref{YeastRobustness}). Maybe more relevant is the fact that,
under fully random functions (RF) Yeast's regulatory dynamics is
chaotic for a wide range of $p$ values. In contrast, the nested
canalizing functions ensure that the dynamics is stable, although it
may be functioning close to the order-chaos boundary if one
effectively has $p \simeq 1/2$.

\begin{figure}[h]
\vspace*{1.0cm}
\begin{center}
\includegraphics[width=5in, angle = 0]{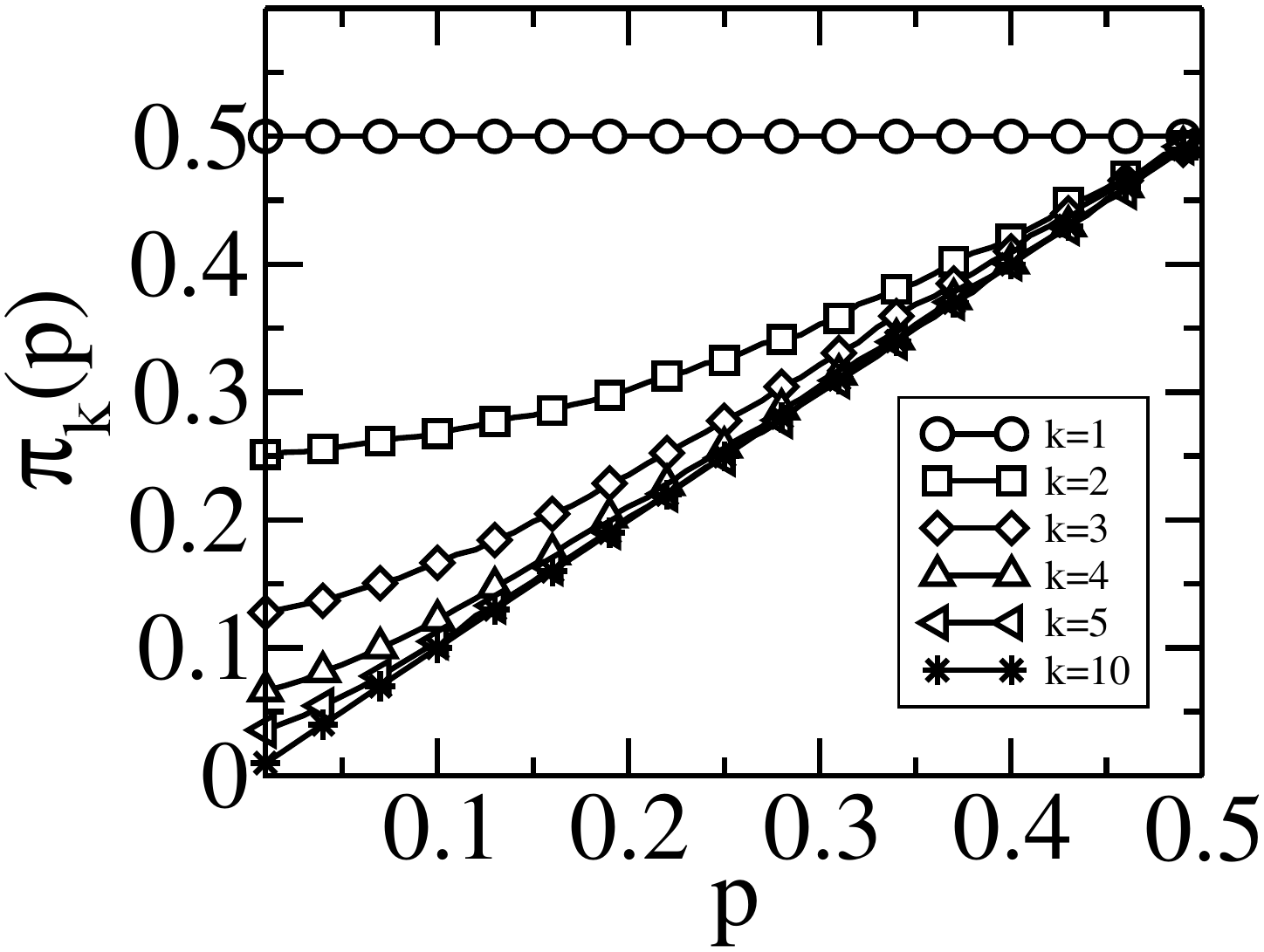}
\end{center}
\caption{}
\label{p_of_k}
\end{figure}

It is worthwhile pointing out that, results in Fig.~\ref{YeastRobustness}
differ from the approximate analytical form formulated by
\cite{Aldana_BooleanNetPLtopology2003} and \cite{Derrida}:
\begin{equation}
\label{r_anal}
r = 2\langle k \rangle p(1-p)\ ,
\end{equation}
where $\langle k\rangle$ is the average in-degree of the network. The
disagreement is due to the condition given in Eq.~(\ref{relevant}). This
experimentally imposed constraint renders the system relatively less robust
with respect to an unbiased choice of the update functions, because functions
filtered by Eq.~(\ref{relevant}) are insensitive to at least one of the input
variables. As a result, the range of the robustness measure $r$ is squeezed
into a narrower interval around the order-chaos boundary $r=1$, conforming to
the edge-of-chaos hypothesis of \cite{Kauffman_Origins_of_Order}.
Furthermore, as seen in Eq.~(\ref{nondegenerate}), the constraint in
Eq.~(\ref{relevant}) is not equally restrictive on all nodes. For example, a
node with $k=1$ has $\pi_1=1/2$ independent of input value $p$, whereas for
$k\gtrsim 5$ we have $\pi_{k}\simeq p$. Generalizing Eq.~(\ref{r_anal})
accordingly, we obtain
\begin{eqnarray}
\label{new_r_anal}
r(p) &=& \sum_{k} 2k\,n(k)\,\pi_k(p)[1-\pi_k(p)]\ ,
\end{eqnarray}
where $n(k)$ is the fraction of nodes with in-degree $k$. $\pi_k(p)$
on the RF class is shown in Fig.~\ref{p_of_k}. The CF and NCF
classes give qualitatively similar behavior. The prediction of
Eq.~(\ref{new_r_anal}) on model networks with the RF-type functions is
also plotted in Fig.~\ref{YeastRobustness} for comparison.

Finally, lifting the constraint of Eq.~(\ref{relevant}) recovers the expected
trend in Eq.~(\ref{r_anal}) for Yeast with RF assignments~\citep{Tugrul_AK},
while the relative order of robustness among different function types is
preserved (not shown).

\subsection{Motifs}
\label{motifs_sec}
Regulatory networks are known to exhibit an abundance of certain
subgraphs~\citep{Milo_etal_Motifs}. \cite{Prill_Motif} list the high
frequency subgraphs for several organisms by comparing the TRN with an
ensemble obtained by reshuffling the edges while keeping the {\it
  total degree} of each node fixed. They establish a connection
between the network structure and its dynamics through the stability
properties of each motif. In particular, the motifs (b),(f) and (g) in
Table~\ref{motifs} were found relatively abundant (with a high
z-score) in Yeast.

A similar investigation comparing Yeast's (unpruned) TRN with the two
reference ensembles E1 \& E2 was reported earlier in Supplement 1 of
\citep{BKME} and is worth re-examining here. Interestingly, the
relatively high content of the motif (b) in Table~\ref{motifs}
reported by \citep{Prill_Motif} is found to be reproduced not only by
E1 but also by the E2 ensemble (the relative frequency of the motif's
occurance is $96.4\%$ in the {\it unpruned} Yeast and E1 networks and
$97.0\%$ in E2). Therefore, this feature should be associated with the
basic matching mechanism of the transcriptional regulation process
exploited in E2, and its high occurance rate in Yeast is guaranteed by
the in- and out-degree statistics (hence the agreement with E1) that
is encoded into the transcription factor binding sequence
statistics. 

On the other hand, cascaded regulation motif (a) and the feed-forward
loop motif (f) of Table~\ref{motifs} appear respectively $\sim 50\%$
and $\sim 80\%$ more frequently in the unpruned Yeast network relative
to E2. The excess of these two motifs in Yeast therefore require a
different explanation, such as the stability considerations given
by~\citep{Prill_Motif}.
\begin{table}[h]
\begin{center}
\includegraphics[width=4in, angle = 0]{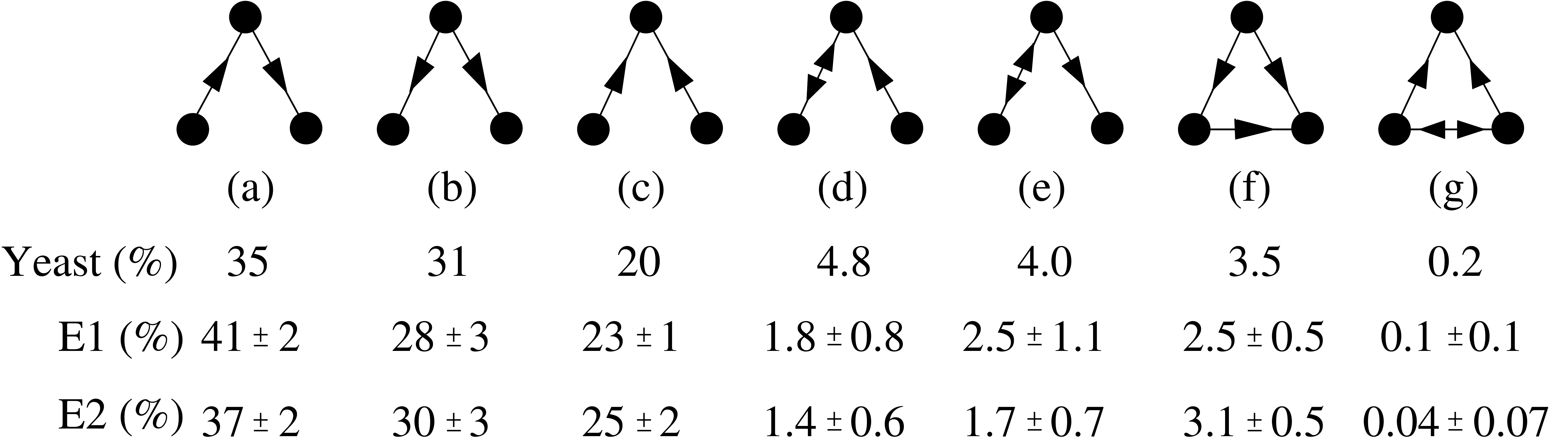}
\end{center}
\caption{}
\label{motifs}
\end{table}

The discrepancy in the attractor statistics (or the lack of it in
respective robustnesses) observed between Yeast and the reference
model DRCs are likely to be related to their motif statistics. On the
other hand, there is no a priori reason that the motif frequencies on
the unpruned network quoted above should also apply to the
DRC. Therefore we performed a similar analysis on the core networks of
Yeast, E1 and E2 using {\it Mfinder}, the free motif-finder software
from Alon Lab~\citep{mfinder}. The results are shown in
Table~\ref{motifs}.

Interestingly, the strong bias towards an excess of dynamically stable
motifs such as (b) and (f) in Yeast does not persist in the DRC. At
the same time, (d) and (e) in Table~\ref{motifs} involving mutually
regulating TF pairs are 60-350\% more frequent in Yeast in comparison
to reference DRCs. These two motifs were labelled as partially stable
by \citep{Prill_Motif}. Our results on network robustness shown in
Fig.~\ref{YeastRobustness}, where no significant difference can be
seen between Yeast and the reference models under Boolean dynamics,
are consistent with such motif statistics.

\section{SUMMARY and DISCUSSION}
We have investigated the dynamical properties of Yeast's
transcriptional regulatory network by means of Boolean functions with
parallel (synchronous) update rules and compared them with two
null-models that capture many of the global structural features found
in Yeast. We found that, the core of the Yeast network (DRC) that
determines the steady-state features of the dynamics is much larger
than the unbiased model (E2) whose sole premise is that the regulation
mechanism is based on sequence-specific binding of transcription
factors. 

Comparison of the average number of attractors (properly weighted by
the basin size) reveals not only the well-known abundance of many
self-regulating genes, but also further differences between the Yeast
and model networks. In particular, we find that the architecture of
the Yeast DRC typically permits several attractors, whereas the model
networks -even after the differences in the number of self-regulating
TFs and the core network sizes are eliminated- come typically with a
single attractor. The tail of the attractor number distribution in
Yeast is also noticably different. An important question is whether
these features survive under a more realistic asynchronous
time-evolution model, although our observations relating to the
network structure remain valid independent of this choice.

A comparison of the network stability under minor perturbations
reveals that the Yeast's dynamical core is {\it not} more robust than
either of the two reference models, in (apparent) contrast with
earlier results. This observation is also supported by the presence of
a different set of 3-node motifs that are found in relatively high
proportions in Yeast when the DRC (rather than the whole regulatory
network) is taken into account. The significantly frequent motifs in
the DRC are dynamically less stable than those found in abundance in the
full TRN of Yeast.

Upon visual inspection, it is not all too unexpected that the stable motifs
(a-c)\&(f) are found mostly on the ``peripherals'' of the network eliminated
by the pruning process. However, it is interesting to observe how topology
itself supports the robustness-responsiveness dichotomy in transcriptional
regulation process. Robustness is associated with the peripheral components
that carry the environmental signals downstream, i.e., small variations are
filtered out before they reach the DRC. On the other hand, DRC itself is
relatively more responsive than the embedding network to changes in the
expression levels of its constituent genes. Favorably so, since such changes
are likely to reflect shifts in operational conditions that are persistent
enough to survive the downstream filtering.

\section*{Acknowledgements}

We would like to thank A. Erzan, M. Mungan and D. Balcan for valuable
discussions that led to the present work and a critical reading of the
manuscript. We appreciate the help of D. Balcan for providing the model
networks at an initial stage. This project was supported by TUBITAK through
the grant TBAG-106T553.

\bibliographystyle{jtbnew}
\bibliography{refs_Tugrul}

\newpage
\noindent{\Large CAPTIONS}\\

\noindent Figure 1: {\bf Extraction of DRC.} A depiction of the pruning procedure used for extracting the
     dynamically relevant core of Yeast. Numbers refer to the stage
     of the recursive process at which the nodes/edges are removed. The red
     edges are the interactions left in the DRC.\\

\noindent Figure 2: {\bf DRC of Yeast.} The dynamically relevant subnetwork of the Yeast's TRN obtained by
      the pruning procedure described above.\\

\noindent Figure 3: {\bf DRC sizes: models {\it vs.} Yeast.} The probability distribution function of the number of genes in the
     DRC with (red and blue) and without (green and black) self-regulating genes.\\

\noindent Figure 4: {\bf Bare {\it vs.} weighted attractor count.} The difference
    between the probability distribution functions for $N_{att}$ (full)
    obtained by plain counting and $\tilde{N}_{att}$ (empty) defined in the
    text reflects the network's preference towards an uneven basin-size
    distribution. (By Eq.~(\ref{entropy}), $\tilde{N}_{att} = N_{att}$ if
    attactors have equal basin sizes.) Presented data is obtained from the E2
    ensemble only for reasons discussed in the text.\\

\noindent Figure 5: {\bf Number of attractors: models {\it vs.} Yeast.} Probability distribution functions for the number of attractors in
    Yeast ($\circ$) {\it vs.} model ensembles with update rules of type RF
    after the self-loops are removed. Data obtained from E1 ($\Box$) and E2
    ($\Diamond$) ensembles are shown together with those over a subset of the
    E2 networks whose DRC sizes are in the same ballpark as that of Yeast's
    ($\bigtriangleup$). Inset shows the tails of the distibutions on a log-log
    scale. Unit size bins were used for the histograms.\\

\noindent Figure 6: {\bf Robustness of Yeast and model networks.} Robustness
of Yeast {\it vs.} model networks with equal DRCs. Horizontal
  axis spans $0\le p \le 0.5$, since $r(p)$ in Eq.~(\ref{r_eq}) is symmetric with
  respect to $p=0.5$. Shaded regions indicate the range within one standard
  deviation of E1 (red) and E2 (blue) reference results. Solid curves
  represent Yeast. Horizontal dashed line is the border between ordered and
  chaotic behavior. Dotted curve corresponds to the theoretical prediction for
  the RF case obtained from Eq.~(\ref{new_r_anal}).\\

\noindent Figure 7: {\bf Impact of the ``relevance'' condition.} The unconditional
    probability $\pi_k(p)$ for a node with in-degree $k$ to be 1 obtained when
    the Random Boolean functions (RF) with bias $p$ are subjected to the
    constraint in Eq.~(\ref{relevant}). These functions are fed into
    Eq.~(\ref{new_r_anal}) in order to obtained an analytical estimate for the
    network robustness. The larger the deviation from the diagonal the
    stronger is the impact of the constraint.\\

\noindent Table 1: {\bf Average attractor number.} Average attractor number obtained for each function class on the DRC
   of the Yeast and the two reference ensembles. E2$^*$ refers to the E2
   ensemble networks which have the same DRC size as Yeast.\\

\noindent Table 2: {\bf 3-node Motif statistics.} Top 3-node motifs ordered according to their frequency of
   appearance in Yeast DRC. Lower percentages are those obtained from
   model networks with the same DRC size as Yeast's.

\end{document}